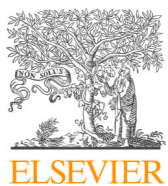
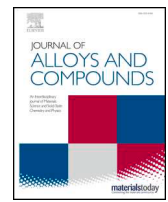

# Dense strontium hexaferrite-based permanent magnet composites assisted by cold sintering process

Eduardo García-Martín[a,b], Cecilia Granados-Miralles[a], Sandra Ruiz-Gómez[b,c], Lucas Pérez[b], Adolfo del Campo[a], Jesús Carlos Guzmán-Mínguez[a], César de Julián Fernández[d], Adrián Quesada[a], José F. Fernández[a], Aida Serrano[a,*,1]

[a] *Departamento de Electrocerámica, Instituto de Cerámica y Vidrio (ICV), CSIC, 28049 Madrid, Spain*
[b] *Departamento de Física de Materiales, Universidad Complutense de Madrid, 28040 Madrid, Spain*
[c] *Max Planck Institute for Chemical Physics of Solids, 01187 Dresden, Germany*
[d] *INEM-CNR, Parco Area delle Scienze 37/A, I-43124 Parma, Italy*

## ABSTRACT

The use of rare-earth-based permanent magnets is one of the critical points for the development of the current technology. On the one hand, industry of the rare-earths is highly polluting due to the negative environmental impact of their extraction and, on the other hand, the sector is potentially dependent on China. Therefore, investigation is required both in the development of rare-earth-free permanent magnets and in sintering processes that enable their greener fabrication with attractive magnetic properties at a more competitive price. This work presents the use of a cold sintering process (CSP) followed by a post-annealing at 1100 °C as a new way to sinter composite permanent magnets based on strontium ferrite (SFO). Composites that incorporate a percentage ≤ 10% of an additional magnetic phase have been prepared and the morphological, structural and magnetic properties have been evaluated after each stage of the process. CSP induces a phase transformation of SFO in the composites, which is partially recovered by the post-thermal treatment improving the relative density to 92% and the magnetic response of the final magnets with a coercivity of up to 3.0 kOe. Control of the magnetic properties is possible through the composition and the grain size in the sintered magnets. These attractive results show the potential of the sintering approach as an alternative to develop modern rare-earth-free composite permanent magnets.

## 1. Introduction

Permanent magnets are crucial materials for the operation of engines and basic elements in turbines used, for example, for the power generation in wind, hydro or thermal power plants. The constant technological demand for these materials leads to an overall production of around 500,000 tons per year, with a high tendency to continue increasing [1]. Currently, the production of permanent magnets requires high percentages of rare-earths (mainly, Nd, Dy, Sm), which make them critical materials for the European Union [2]. In addition, the extraction of rare-earths produces an extremely negative environmental impact. Therefore, the development of new permanent magnets that preserve or exceed the high performance of the current ones while being free of rare-earths in their composition is imperative for real applications [3–5], representing a great challenge.

As base of rare-earth-free permanent magnets, the hexaferrites ($MFe_{12}O_{19}$, M = Pb, Sr, Ba) are very interesting materials which can be also used in a wide range of applications such as magnetic recording and data storage materials as well as in electrical devices that operate at microwave/GHz frequencies [6]. Specifically, hexaferrites can be employed for the manufacturing of devices that must work under certain environmental conditions, due to their high coercivity (magnetic hardness) and high resistance to corrosion [1,6–12]. A possible way to improve the properties of rare-earth-free permanent magnets is to generate biphasic materials composed of a magnetically hard material, for instance hexaferrites (which provides high coercivity) and a softer material (which provides high magnetization) [1,13,14], improving the properties with respect to the hard material, with high magnetization and additional hardening. Another approach consists of avoiding the grain coarsening process of material but ensuring the densification process of pieces [15,16]. For that, the incorporation of secondary phases in the hard matrix is the most widely used procedure [17,18]. However, high temperatures above 1200 °C are usually required and other interesting routes at lower temperatures and sintering times may be considered in order to reduce the energetic costs, the environmental impacts and allowing an opportunity for multiphase composites to survive the densification stage.

* Corresponding author.
*E-mail address:* aida.serrano@icv.csic.es (A. Serrano).
[1] ORCID: 0000-0002-6162-0014.

In 2016, Randall et al. presented for the first time the method of cold sintering process (CSP) as a novel via for the sintering of ceramics at lower temperatures under an applied pressure. This method is based on the combination of at least one inorganic compound in particle form with an aqueous solvent that partially solubilizes the high chemical potential sites of the particle surfaces, which is processed under uniaxial pressure (> 0.5 bar) and low temperatures (< 300 °C), in order to form a sintered material [19–22]. This processing method has been used to prepare a large number of materials and compounds [19,21–24]. In particular, we have recently reported the first CSP rare-earth-free permanent magnet, based on hexaferrites, using novelty glacial acetic acid as solvent [25] and we have researched the effect of the organic solvent on the CSP [26], with very attractive and industrially scalable results. To date, some further research has been carried out on CSP-assisted hexaferrites. For example, Lowum et al. have sintered $BaFe_{12}O_{19}$ by hydroflux-assisted densification technique based on CSP [27] and Rajan et al. have densified $Li_2MoO_4$–$SrFe_{12}O_{19}$ composites by CSP using water as solvent [28]. However, no studies have been yet performed on CSP of hexaferrite-based composites as permanent magnets.

In this work, composite permanent magnets using strontium hexaferrite ($SrFe_{12}O_{19}$, SFO) have been prepared incorporating several secondary phases in the nano and micrometric scale: Fe-Si, $Fe_3O_4$, Fe-$Fe_3O_4$ and SFO. Secondary phases based on Fe and $Fe_3O_4$ could help to increase the saturation magnetization while the incorporation of SFO with smaller particles could promote greater densification of the final magnets. A novel approach consisting on a CSP intermediate step using glacial acetic acid as solvent plus a post-thermal annealing at 1100 °C for 2 h is applied for the first time to composite permanent magnets. Additionally, characteristic morphological, compositional and structural properties are identified depending on the starting secondary phase added and the sintering step of composite magnets, influencing in their final magnetic response.

## 2. Experimental methods

### 2.1. Fabrication of samples

SFO powders from Max Baermann Holding (Germany) [29], with a platelet-like morphology and a bimodal particle size of 100–500 nm and 1–3 µm (see Supporting Information, SI), were combined with a secondary magnetic material to sinter SFO-based composite ceramic permanent magnets. The secondary magnetic material consisted of: i) 10 wt% of SFO nanoparticles (nSFO) synthetized following the method described by Guzmán-Mínguez et al. in [16] with an average size of 100 nm, ii) 10 wt% of Fe-3.5%Si alloy microparticles (Fe-Si) from Thyssenkrupp (Germany) [30] with a micrometric bimodal distribution of an average size of 3 µm and 12 µm, iii) 5 wt% of nanofer star particles from Nanoiron company (Czech Republic) [31] with an average size about 100 nm and a $Fe_3O_4$ wt% of 13%, forming Fe-$Fe_3O_4$ core-shell structures (Fe-$Fe_3O_4$) and iv) 10 wt% of commercial $Fe_3O_4$ particles ($Fe_3O_4$) with and bimodal particle size of 400 nm and 1.5 µm. The percentage of each secondary phase was determined to generate final pieces with the best densification. A morphological and compositional characterization of different starting particles is shown in SI.

**Table 1**
Description of samples sintered for this work after the intermediate CSP step ($P_{CSP}$ = 2.5 bar; $T_{CSP}$ = 190 °C; solvent: glacial acetic acid) and after the final post-annealing step at 1100 °C for 2 h (marked with an *), indicating the initial compositions and the sintering story of the samples.

| Sample | Initial composition (wt%) | Sintering process |
| --- | --- | --- |
| S1 | 90SFO + 10nSFO | CSP |
| S1* | 90SFO + 10nSFO | CSP +post-annealing |
| S2 | 90SFO + 10$Fe_3O_4$ | CSP |
| S2* | 90SFO + 10$Fe_3O_4$ | CSP +post-annealing |
| S3 | 95SFO + 5Fe-$Fe_3O_4$ | CSP |
| S3* | 95SFO + 5Fe-$Fe_3O_4$ | CSP +post-annealing |
| S4 | 90SFO + 10Fe-Si | CSP |
| S4* | 90SFO + 10Fe-Si | CSP +post-annealing |

The mixture of SFO powders with nSFO, Fe-Si, Fe-$Fe_3O_4$ and $Fe_3O_4$ particles was prepared by a 10 min-long dry dispersion method [32,33] and the final product in each case was submitted to sintering process, as previously described in [25]. Initially, the SFO-based powder mixture was mixed with glacial acetic acid as CSP solvent in a ratio of 50:50 wt% and then, the obtained granulated powders were pressed at 1 bar for 5 min at room temperature in a cylindrical die with an inner diameter of 0.83 mm. Subsequently, the pressed pellets of all mixtures were submitted to CSP in a press BURKLE D-7290 heating at 190 °C for 2 h with an annealing rate of 20 °C/min while applying a pressure of 2.5 bar, and then cooled down in air. CSP parameters were previously optimized looking for the highest density value of pieces. Finally, SFO-based composites obtained from CSP were annealed at 1100 °C for 2 h in air atmosphere as final step, with an annealing rate of 5 °C/min. Post-annealing conditions were chosen according to previous work [25,26]. Final composite ceramic pieces and those obtained from the intermediate CSP step are selected and labelled in Table 1.

### 2.2. Characterization of samples

The sample density was calculated by averaged values obtained by two different methodologies: (i) the measurements of mass/dimensions and (ii) a specific procedure for high density determination by hydrostatic weighing for porous samples based on the Archimedes method using water as liquid medium [34]. The relative density values were determined considering the theoretical density determined as a weighted average of 5.10 g/cm$^3$ for SFO [35], 5.26 g/cm$^3$ for $\alpha$-$Fe_2O_3$ [36] and 5.20 g/cm$^3$ for $Fe_3O_4$ [37], and considering the compositional phases obtained from Rietveld analysis of the X-ray diffraction (XRD) data after the sintering process.

The grain morphology in the composite ceramics based on SFO was studied by field emission scanning electron microscopy (FESEM) with an S-4700 Hitachi instrument at 20 kV on fresh fractured surfaces for all samples. The crystalline fraction of the samples was studied by powder XRD. XRD data were collected using a D8 diffractometer from Bruker, equipped with a Lynx Eye detector and a Cu K$\alpha$ target ($\lambda$ = 1.5406 Å). Rietveld analysis of the XRD data was carried out using the software FullProf [38]. Quantitative information on the composition, the average crystalline size and the lattice dimensions was extracted from the Rietveld refinements. Specifications on the Rietveld model fitted to the data may be found in the SI.

Confocal Raman microscopy measurements were performed using a Witec alpha-300RA instrument with a Nd:YAG laser of 532 nm. Raman experiments were carried out by fixing the laser excitation power at 0.5 mW and using an objective with a numerical aperture (NA) of 0.95. An average Raman spectrum was obtained for each sample from an in-plane mapping of 10 × 10 µm$^2$ on the sample surfaces, where each single Raman spectrum is recorded every 200 nm. Raman results were analyzed by using Witec Project Plus Software.

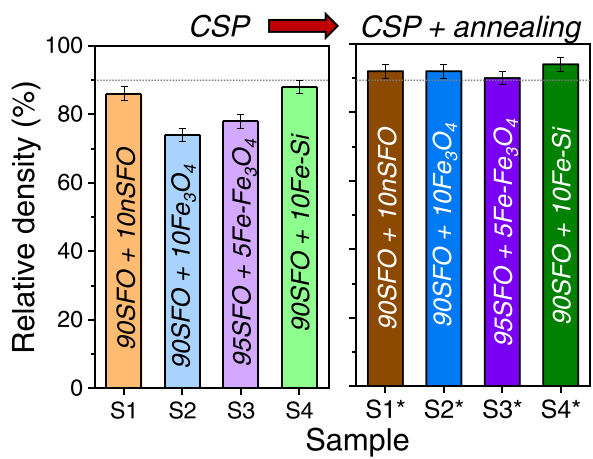

**Fig. 1.** Evolution of relative density for SFO-based composite magnets processed by CSP using glacial acetic acid and after an annealing step at 1100 °C for 2 h of different SFO-based compositions. Relative density is calculated with respect to the theoretical density value considering 5.10 g/cm$^3$ for SFO [35], 5.26 g/cm$^3$ for α-Fe$_2$O$_3$ [36] and 5.20 g/cm$^3$ for Fe$_3$O$_4$ [37], and taking into account the compositions extracted from Rietveld refinements of the XRD data. The final relative density values are higher than 90% for all post-annealed composites. Grey dashed line denotes 90% of relative density.

Finally, to determinate the magnetic properties, ceramic magnets were measured in a vibrating sample magnetometer (VSM) Lakeshore 7304. The magnetic signal is measured at room temperature applying a magnetic field up to 11 kOe. The sample (cylindrical pellet) was mounted so that the direction of the applied magnetic field was perpendicular to the pellet surface, i.e., parallel to the direction of the uniaxial pressure applied to fabricate the pellets. It should be noted that maximum field applied is 11 kOe, so the samples are not fully saturated. Therefore, here the magnetization values are given at 11 kOe ($M_{11\ kOe}$) instead of saturation magnetization ($M_S$) [15].

## 3. Results and discussion

### 3.1. Morphological and structural characterization

Fig. 1 shows the relative density of the ceramic composites prepared by the sintering process, after the CSP step and after the post-thermal treatment at 1100 °C for 2 h. Significant modifications depending on the incorporated secondary phase powders are found. After the intermediate CSP step, the lowest relative density, around 74%, is obtained for S2 sample in which a 10 wt% of Fe$_3$O$_4$ particles is incorporated in the composition along with the SFO. This value is even lower than that previously reported for the same single hexaferrite powders sintered by same CSP [25]. The largest relative density values around 88% are attained for S1 and S4 hexaferrite-based ceramics with nSFO and Fe-Si particles, respectively. These two show sufficient mechanical integrity to be considered as dense ceramic pieces (>85%), being sintered without reaching high temperatures which are usually required in the traditional thermal sintering processes (>1200 °C) [18]. In these compositions, the glacial acetic acid facilitates the mass transport and the occurrence of the dissolution-precipitation processes, followed by the organic solvent evaporation that allows the densification [20,33]. However, there is no common pattern in the densification by CSP since the best densities are obtained when the second phase is SFO nanoparticles or metallic microparticles. Given that the partial solubility of SFO in glacial acetic acid has been shown to be the main factor in densification by CSP [25], the use of SFO nanoparticles contributes to such cold densification due to their larger surface area. Additionally, chemical attack in an acid medium of the metallic particles also contributes to cold densification. On the contrary, Fe$_3$O$_4$ particles seem to present a lower dissolution in glacial acetic acid medium. In the case of Fe-Fe$_3$O$_4$, the presence of a protective passivation layer consisting of Fe$_3$O$_4$ [31], would justify the low contribution of this second phase to the cold densification of the composites.

When CSP samples are annealed at 1100 °C after the CSP step, an increase of the relative density is distinguished for all cases, reaching values around 92% with respect to the theoretical density value calculated for each sample considering the different phases identified below by XRD technique. The post-annealing process at higher temperature on the CSP pieces involves the mass transport, Ostwald ripening and recrystallisation phenomena, densifying the magnetic pieces. Density results were similar independently of the secondary phase incorporated initially to SFO powders. This densification is suitable for sintered ceramic composites that can be employed as ceramic magnets in current magnetic applications. It is noteworthy that these density values could not be reached when the composites are sintered at 1100 °C for 2 h without the CSP step. Relative densities below 80% are obtained by conventional sintering at 1100 °C (not shown), as for single SFO sintered under same temperature [25]. Higher temperatures than 1200 °C are usually required to promote the densification process of SFO-based magnets, as commented above, indicating the important role of the CSP stage in densification process.

The effect of the sintering process on the ceramic morphology of different composites based on SFO is displayed in Fig. 2. The glacial acetic acid used as solvent in the CSP step enables local dissolution of the particle surface, and a great integration of the different particles dispersed in the SFO matrix is observed, identifying a large homogeneity of compounds with low open porosity (see Fig. 2a–d). Specifically, S1 sample presents the lowest proportion of particles with an average grain size lower than 400 nm, likely due to the coalescence of SFO nanoparticles or their integration with larger grains. Composites prepared from Fe$_3$O$_4$, Fe-Fe$_3$O$_4$ and Fe-Si particles consist of the largest quantity of nanograins (Fig. 2b–d and Fig. S4 in SI) and present a clear bimodal grain size distribution with a greater or lesser fraction of each depending on initial secondary powders and their proportion. In any instance, the average grain size in all the composites is similar with a value of around 0.7 µm (see grain size distribution analysis in Fig. 2).

After the post-annealing step at 1100 °C for 2 h of the CSP composites, a clear change in the grain morphology and the average grain size is found. For the S1* sample, more rounded and homogeneous grains are noted with a value around 0.6 ± 0.4 µm, similar on average than the S1 sample (previous step). A large quantity of small grains is still identified after the final sintering process (see Fig. 2e). Conversely, for the rest of composites (S2*, S3* and S4* samples), the coalescence of smaller grains in equiaxed grains during the solid-state sintering is recognized (Fig. 2f-h). The grain morphology tend to acquire a form of platelets. The effect is quite similar independently on the initial secondary phase. A grain growth is identified, leading to grains slightly larger than 1 µm, with straight grain boundaries and platelets interconnected between them. The presence of straight grain boundaries indicates that the equilibrium in the sintering driving forces has been attached during the post-annealing process of the CSP composites. This fact clearly indicated the diffusion activation of the chemical species to attain a final step of sintering at temperatures lower than the usually needed for sintering ceramic ferrites based on SFO [15]. It is noteworthy that S1* sample shows a morphology quite different to the rest of sintered samples and that obtained by conventional route [25], probably due to the nature of the secondary phase that makes the glacial acetic acid solvent act differently during the CSP process. On one hand, as commented above, the CSP step (S1 sample) induces a morphology with lest number of nanograins than in the other samples. On the other hand, this sample is the one constituted by SFO without secondary phases (see XRD and Raman results), indicating that the behavior during the sintering process is different than in the other magnets. Therefore, the influence of the secondary phases and the sintering steps in the route is distinguished in the morphological characteristics of the samples.

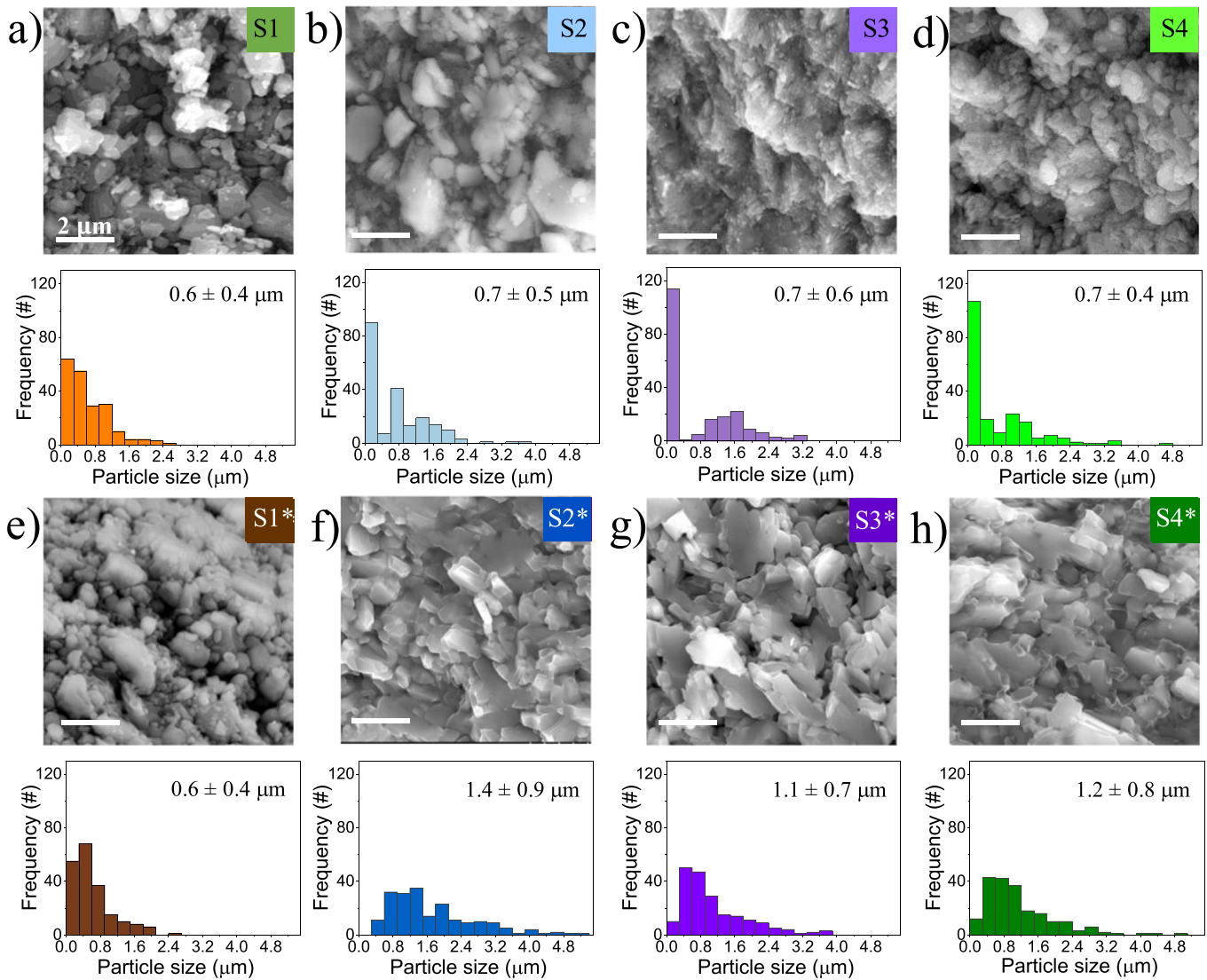

**Fig. 2.** FESEM micrographs and grain size distribution analysis of composite ceramics processed after the CSP step: a) S1, b) S2, c) S3 and d) S4 and these post-annealed at 1100 °C for 2 h: e) S1*, f) S2*, g) S3* and g) S4*. The white bars correspond to 2 μm.

Powder diffraction gave information on the crystalline phases present in the various samples. Refined weight fractions for each phase, retrieved from Rietveld refinements of the XRD data, are represented in Fig. 3a-b. Exact weight fractions from each phase and sample may be found in the SI. A partial decomposition of the ferrite phase is induced by the CSP step under pressure at low temperature (190 °C). During the process, the glacial acetic acid partially solubilizes the highest chemical potential regions of the particles, facilitating their rearrangement, interdiffusion and promoting mass transport. The densification process is given by the incongruent dissolution, which leaves the phase transformation, and the re-precipitation process, as was previously identified in single SFO particles [25]. The magnitude of the phase transformation varies depending on type and size of particles are incorporated at SFO matrix in the starting mixture. Thus, the largest quantity of SFO after the CSP is identified in S1 sample, which only contains a 4% of α-$Fe_2O_3$ with the rest as SFO, while the largest transformation is obtained for S4 sample with a 97% of $Fe_3O_4$ and just a 3% of SFO phase.

S2 and S3 samples exhibit a little more than 50% of SFO along with a mixture of α-$Fe_2O_3$ and $Fe_3O_4$ polymorphs. The smallest transformation of SFO is obtained in S1 sample, where nanoparticles of the same phase (SFO) were incorporated initially, which must be inducing a stabilization to the system and in turn, reducing the phase decomposition during the CSP. However, the presence of microparticles of Fe-3.5%Si alloy in the S4 sample leaves a large solubilization of the particles induced by the glacial acetic acid with the increased SFO decomposition, which may be related to the nature and/or the micrometer size of the secondary phase.

After the post-thermal treatment at 1100 °C of the CSP pieces, part of the SFO phase is recovered achieving a 100%, 87%, 95% and 76% for S1*, S2*, S3* and S4* sample, respectively, along with α-$Fe_2O_3$ as the only secondary phase in all cases, the most thermodynamically stable iron oxide phase. Concerning the lattice dimensions and volume-averaged crystallite sizes for SFO phase, no trend is identified with respect to the type of incorporated secondary material or the sintering steps followed during the process (see SI).

In order to confirm the crystalline phases distinguished by XRD in the samples and to identify possible additional phases with low concentration (not previously detected by XRD) or/and amorphous compounds, an analysis by confocal Raman microscopy has been carried out. In addition, a Fourier self-deconvolution (FSD) is performed in the experimental average Raman spectra as a mathematical tool to differentiate the vibrational modes. The purpose of FSD is to narrow the spectral bands, preserving the frequency and the integrated intensity of each line [39]. In order to reach the best compromise between spectral resolution and the deconvolution of the Raman modes the parameters gamma and smoothing factor were fixed to 10 and 0.2, respectively. Fig. 4 shows the average Raman spectra obtained for different ceramic composites sintered after the CSP step and after the post-annealing step, as well as the spectra from the FSD performed in the experimental averaged Raman spectra.

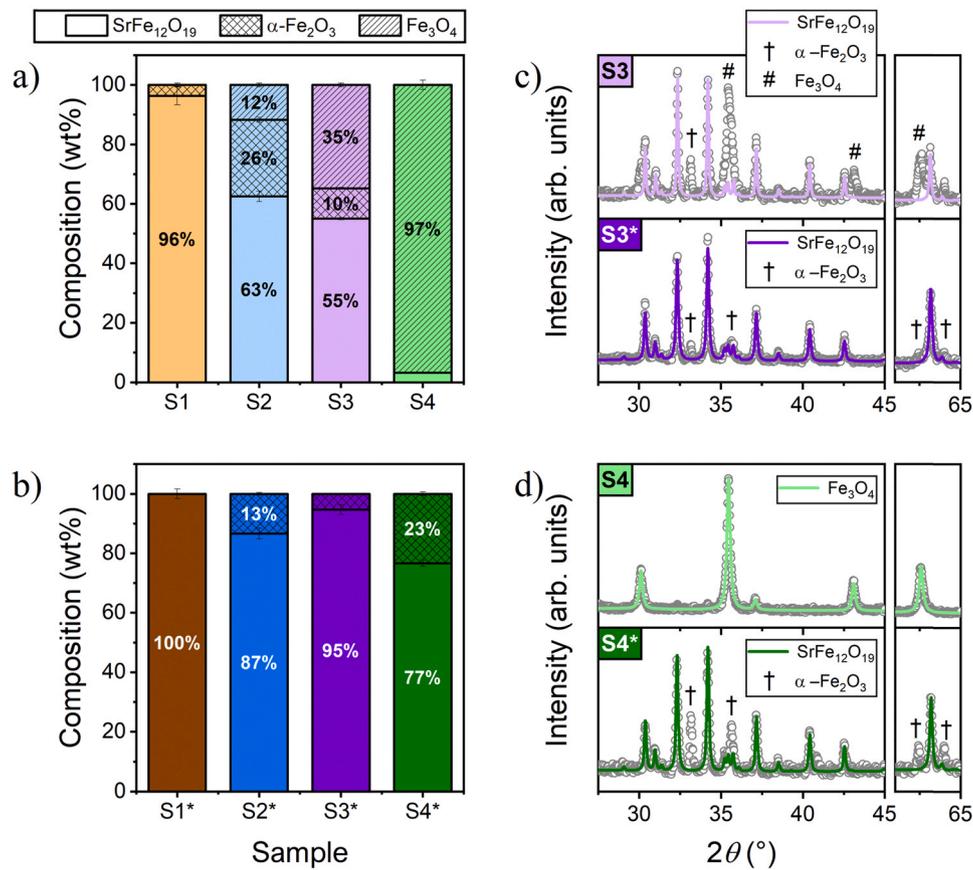

**Fig. 3.** a-b) Phase composition in weight fraction extracted from Rietveld refinements of the XRD data measured for the sintered ceramics processed by CSP (S1–S4) and by CSP followed by a post-annealing at 1100 °C for 2 h (S1*-S4*). Percentages smaller than 10% are not labelled in the Figure. c-d) For selected samples (S3, S3*, S4, and S4*), measured XRD data (grey open circles), Rietveld model built for the main phase in each case (solid line in colour) and Bragg positions of the minority phases (†, #).

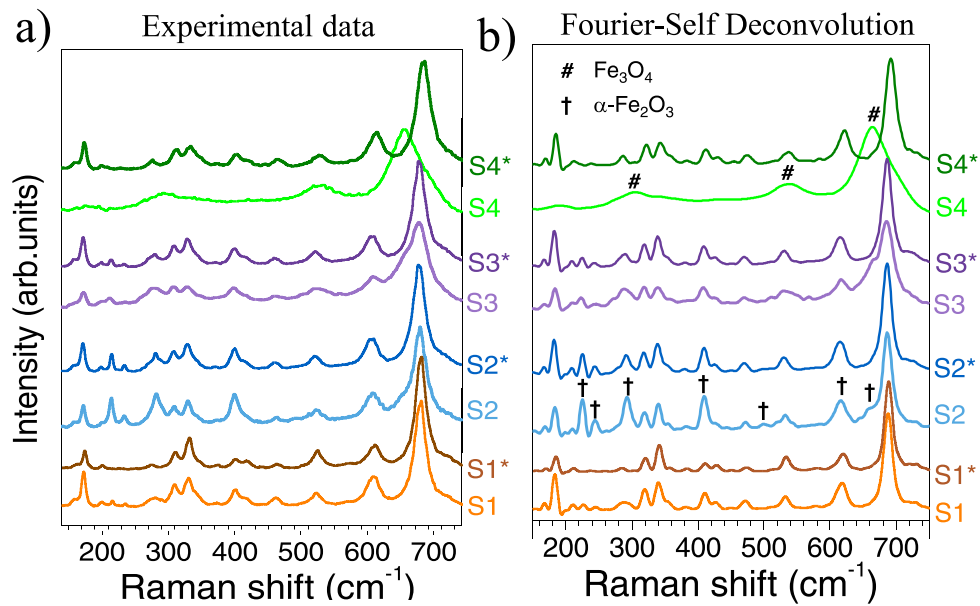

**Fig. 4.** a) Raman and b) FSD spectra of the ferrite-based composited sintered after the CSP step and plus a post-annealing at 1100 °C for 2 h (samples marked with *). Spectra are presented from 150 to 750 cm$^{-1}$. Raman modes related to hematite and magnetite phases (†, #) are identified on the Figure in the Raman spectra with the largest fraction of the corresponding iron oxide polymorph.

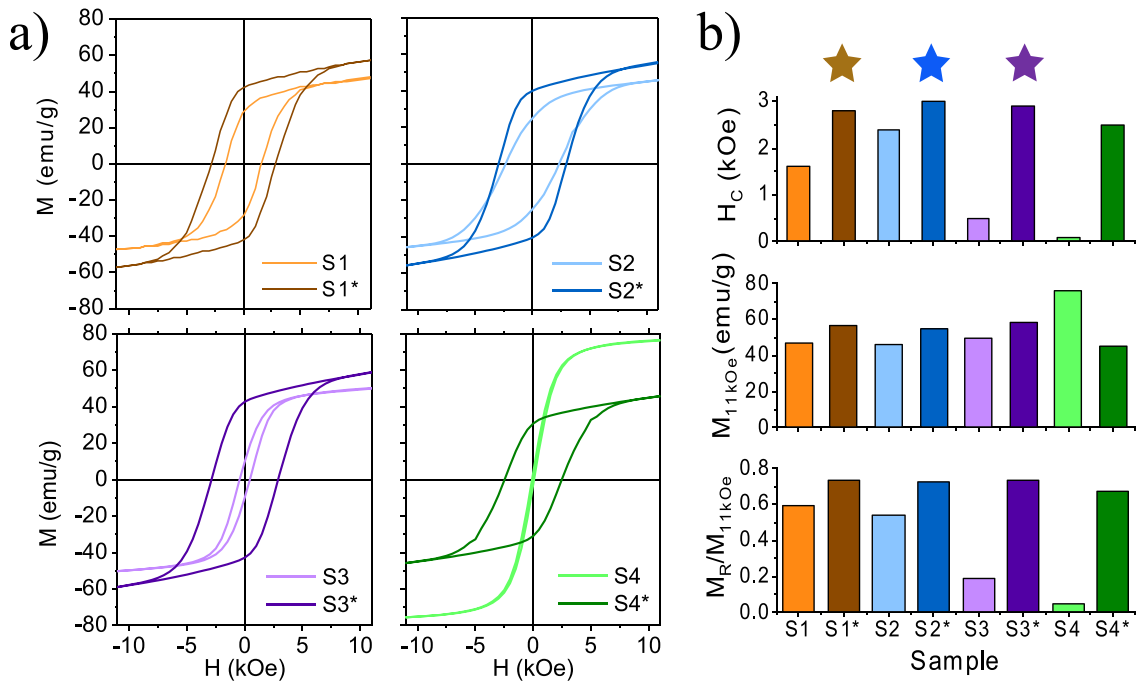

**Fig. 5.** a) Hysteresis loops displaying the magnetic response of hexaferrite-based composites after the CSP step: S1, S2, S3 and S4 and these after a post-annealing step: S1*, S2*, S3* and S4*. The plotted loops have not been corrected for demagnetization. b) Magnetic characteristics obtained from the hysteresis loops for each sample: $H_C$, $M_{11\,kOe}$ and $M_R/M_{11\,kOe}$, indicating the best ones with a star.

According to the group theory of hexaferrites based on the $D_{6h}$ symmetry, 42 Raman modes ($11A_{1g} + 14\,E_{1g} + 17\,E_{2g}$) are active [40–42]. Some of these vibrational bands corresponding to the SFO are identified in most of the composites processed by CSP, in addition to other Raman modes related to secondary $Fe_3O_4$ and $\alpha$-$Fe_2O_3$ phases [43–45], as shown in Fig. 4. Particularly, S4 sample only exhibits the Raman vibrational modes related to the $Fe_3O_4$ phase, as the majority phase recognized by XRD analysis. After the post-annealing process of the CSP pellets, the SFO phase is partially recovered, recognizing some vibrational bands attributed to $\alpha$-$Fe_2O_3$ polymorph in all ceramics except for S1* sample where pure SFO is obtained (see Fig. 4). An assignment of the Raman bands is made taking into account previous works [40–45], and identified vibration modes are presented in Table S4 in the SI. Raman modes corresponding to other phases are not distinguished after the CSP step and post-annealing process, in agreement with the above XRD results.

Additionally, modifications in the position, full width high maximum (FWHM) and intensity of the Raman bands are found. The differences in the relative intensity between the SFO bands is attributed to different orientations of the hexaferrite platelets [40,42], while the intensity of Raman bands associated with the $Fe_3O_4$ and $\alpha$-$Fe_2O_3$ phases with respect to that of the SFO bands could be explained by a greater or lesser proportion of the iron oxide polymorphs (see Fig. 4). A quantification of phases contained in the samples from the Raman data is not possible due to the influence of the orientation of the particles on the intensity of the vibrational bands. The variations in the Raman shift and the FWHM may be associated with changes in the grain size or strain in the structure induced during the sintered process.

At this point it is important to mention that during the decomposition of the SFO phase in the sintering process, no Sr ions were identified outside the SFO structure. In previous works where single SFO was sintered by CSP [7,8], in addition to the crystalline iron oxide phase, amorphous SrO was located by X-ray absorption spectroscopy (XAS) as phase produced during the transformation process of the SFO. The amorphous character of SrO compound did not allow identifying it by XRD technique and was not probably recognized by Raman technique due to its lower scattering cross-section with respect the $SrFe_{12}O_{19}$ and iron oxides presented in samples [46].

### 3.2. Magnetic properties

The magnetic characterization of hexaferrite-based composites both after the CSP and the post-annealing process at 1100 °C is presented in Fig. 5. The typical response of a ferromagnetic material is identified in all cases, with specific properties attributed to the morphological, structural and compositional features of hexaferrite-based ceramics that depend on the starting secondary phase incorporated into composite. When the CSP step is carried out in the different compositions, the hard magnetic properties are reduced as a function of the amount of Sr ferrite transformed into secondary phases such as $\alpha$-$Fe_2O_3$ and $Fe_3O_4$. $\alpha$-$Fe_2O_3$ is a weak canted antiferromagnetic at room temperature and has a low saturation magnetization ($M_s$~1–2 emu/g) while $Fe_3O_4$ exhibits a soft ferrimagnetic behavior produced by the magnetic moments aligned antiparallel to one another resulting in a spontaneous magnetization [47]. Specifically, S2 sample fabricated with 10 wt% of $Fe_3O_4$ particles shows the highest $H_C$ (2.4 kOe), but its relative density is low and its mechanical integrity is poor to be employed in current applications. S1 and S4 samples present the greatest relative density values after the CSP step, larger than 85%, and they may be considered as sintered ceramic pieces. However, while S1 sample shows $H_C$ of 1.6 kOe and $M_{11\,kOe}$ = 47 emu/g with interesting properties to be employed in certain applications, S4 sample presents a soft magnetic response characteristic of its composition with the largest proportion corresponding to $Fe_3O_4$, excluding it as a candidate for permanent magnet material. Therefore, a modulation of the magnetic properties is obtained, related mainly to the phase transformation during CSP step, which depends on the secondary phase incorporated into the SFO matrix.

As shown above, in addition to the partial recuperation of the ferromagnetic phase (SFO), a higher integrity of the pieces is achieved after thermal post-treatment at 1100 °C. With respect to the magnetic properties, all samples except S4* exhibit a $H_C$ ≥ 2.8 kOe (achieving even 3.0 kOe for the S2* magnet) and a $M_{11\,kOe}$ ≥ 55 emu/g. The post-annealed magnet with the worst magnetic properties is S4*, although it still displays attractive properties ($H_C$ = 2.5 kOe and $M_{11\,kOe}$ = 46 emu/g). Comparing the magnetic properties with those obtained in a ceramic magnet processed by the same approach from single SFO particles [25], an improvement of $H_C$ is identified in S2* and S3* samples, prepared using $Fe_3O_4$ and $Fe$-$Fe_3O_4$ as secondary phase, respectively, showing the importance in the nature of secondary phase.

At this point it is relevant to mention that sintered ceramic magnets are competitive with respect to commercial rare-earth-free ferrite magnets (Hitachi metals NMF-7C series offer $H_C$ = 2.8–3.3 kOe and $M_S$ = 68 emu/g) [48], reaching similar performance at lower sintering temperatures (8–12%) than usual thermal procedures of ferrite-based composite permanent magnets at 1200–1250 °C [17,18].

Another feature to be highlighted after the post-annealing process at 1100 °C is the degree of magnetic alignment between particles in the absence of an external magnetic field, which increases in all final ceramic composites processed. A $M_R/M_{11\,kOe}$ ratio ≥ 67% is obtained, even achieving 73% for S1* and S3*, surpassing the predicted value for an assembly of randomly oriented, non-interacting particles [49] and pure SFO powders processed by same methodology [25]. These magnetic properties can be explained by the favourable sliding and rearrangement of the hexaferrite platelets promoted by the glacial acetic acid solvent under pressure during the dissolution-precipitation process in the CSP step, whose values are improved with the mass transport during the post-annealing process.

It is worth noting that the sintering process employed here, consisting on a CSP stage followed by a post-annealing at 1100 °C for 2 h, results in a reduction in the sintering temperature and time which means about 9 kWh/Kg in the energy consumption with respect to that typically required to sinter hexaferrites-based magnets (conventional sintering processes at 1200 °C for 4 h) [50], leading to an energy efficiency of around 29%.

## 4. Conclusions

In this work, rare-earth-free composite permanent magnets based on strontium hexaferrite have been sintered by a process consisting on a CSP intermediate step using glacial acetic acid as solvent plus a post-thermal treatment at 1100 °C for 2 h. The homogeneous incorporation of several secondary phases in the nano and micrometric form has been performed, inducing singular morphological, structural and magnetic features. While the CSP step induces the partial transformation of SFO phase, the post-annealing process partially recovers it, densifying the pieces to a relative density of 92%, controlling the grain size of the final magnets and improving their magnetic response. These results open a new way to develop rare-earth-free composite permanent magnets reducing the energy consumption during the sintering process of ferrite-based composites and with very attractive properties for current applications of ferrite magnets.

## CRediT authorship contribution statement

**Eduardo García-Martín:** Methodology, Formal analysis, Investigation, Data curation, Writing – review & editing. **Cecilia Granados-Miralles:** Formal analysis, Investigation, Data curation, Writing – review & editing. **Sandra Ruiz-Gómez:** Formal analysis, Investigation, Writing – review & editing. **Lucas Pérez:** Validation, Formal analysis, Investigation, Writing – review & editing, Supervision. **Adolfo del Campo:** Investigation, Writing – review & editing. **Jesús Carlos Guzmán-Mínguez:** Investigation, Writing – review & editing. **César de Julián Fernández:** Investigation, Writing – review & editing. **Adrián Quesada:** Conceptualization, Validation, Resources, Writing – review & editing, Funding acquisition. **José F. Fernández:** Conceptualization, Validation, Resources, Writing – review & editing, Funding acquisition. **Aida Serrano:** Conceptualization, Methodology, Validation, Formal analysis, Investigation, Resources, Data curation, Writing – original draft, Writing – review & editing, Supervision, Funding acquisition. All authors have read and agreed to the published version of the manuscript.

## Declaration of Competing Interest

The authors declare that they have no known competing financial interests or personal relationships that could have appeared to influence the work reported in this paper.

## Acknowledgments

This work has been supported by the Ministerio Español de Ciencia e Innovación (MICINN), Spain, through the projects MAT2017-86540-C4-1-R and RTI2018-095303-A-C52, and by the European Commission through Project H2020 No. 720853 (Amphibian). C.G.-M. and A.Q. acknowledge financial support from MICINN through the "Juan de la Cierva" program (FJC2018-035532-I) and the "Ramón y Cajal" contract (RYC-2017-23320). S. R.-G. gratefully acknowledges the financial support of the Alexander von Humboldt foundation, Germany. A.S. acknowledges the financial support from the Comunidad de Madrid, Spain, for an "Atracción de Talento Investigador" contract (No. 2017-t2/IND5395).

## Author contributions

All authors have approved the final version of the manuscript.

## Appendix A. Supporting information

Supplementary data associated with this article can be found in the online version at doi:10.1016/j.jallcom.2022.165531.